\documentclass[prd,aps]{revtex4}

\advance \topmargin by .6in

\begin{document}

\title{ (1+1)-dimensional formalism and 
quasi-local conservation equations}
\author{Jong Hyuk Yoon}
\affiliation{
Department of Physics, Konkuk University, \\
Seoul 143-701, Korea \\
and \\
Enrico Fermi Institute, University of Chicago, \\
5640 S. Ellis Av., Chicago, IL 60637, U.S.A.\\
{\tt yoonjh@gr.uchicago.edu}}

\begin{abstract}
A set of exact quasi-local conservation equations is obtained in the
(1+1)-dimensional description of the Einstein's equations of 
(3+1)-dimensional spacetimes. These equations are interpreted as quasi-local 
energy, linear momentum, and angular momentum conservation equations.
In the asymptotic region of asymptotically flat spacetimes, 
it is shown that these quasi-local conservation equations reduce to the 
conservation equations of Bondi energy, linear momentum, and 
angular momentum, respectively. 
When restricted to the quasi-local horizon of a generic spacetime, 
which is defined without referring to the infinity,  
the quasi-local conservation equations 
coincide with the conservation equations on the stretched horizon
studied by Price and Thorne. All of these quasi-local quantities are 
expressed as invariant two-surface integrals, and geometrical 
interpretations in terms of the area of a given 
two-surface and  a pair of null vector fields orthogonal to 
that surface are given. 
\end{abstract}

\pacs{PACS numbers: 04.20.Cv, 04.20.Fy, 04.20.-q, 04.30.-w;
quasi-local conservation law, black hole dynamics, 
holographic principle, Kaluza-Klein theory}

\maketitle

\begin{section}{Introduction and Kinematics}{\label{intro}}

For the past few decades, there has been enormous progress in 
general relativity since the pioneering works of the late Professor 
A. Lichnerowicz on mathematical relativity. His contributions
to general relativity are diverse as well as profound, not least 
because he put the Einstein's equations on a firm mathematical foundation
as partial differential equations and theories of 
connections\cite{lichnerowicz55}. Connections also plays 
important roles in Yang-Mills gauge theories, since gauge theories are 
nothing but theories of connections coupled to matter fields.
Therefore, in this International Conference commemorating Professor 
A. Lichnerowicz, it seems appropriate to discuss a relatively unknown 
formalism of general relativity, which is based on the very idea of connections.

This note is about (1+1)-dimensional description of general 
relativity of (3+1)-dimensional spacetimes, treating the remaining 
2-dimensional spatial dimensions as a fibre space. 
In this framework, all the notions in the theory of fibre bundles 
such as a fibre space, connections, and the structure group
appear naturally. 
Instead of going into the details of the formalism 
itself\cite{yoon92,yoon93a,yoon01}, however, 
I will describe the key ideas briefly, mainly to fix the notations, 
and then quickly move on to discuss
issues that are more immediate, namely, the problem of defining 
the quasi-local conservation equations using the (1+1)-dimensional 
formalism of (3+1)-dimensional spacetimes.

Let us begin by mentioning a few facts about quasi-local conservation
equations. 
In general relativity there have been many attempts to obtain 
quasi-local conservation 
equations\cite{yoon01,bro-york93,hay94,nest91,sza94,berg92,yoon99c}.
One of the motivations of these efforts is the expectation 
that quasi-local conservation equations allow us
to predict certain aspects of a quasi-local region of a given spacetime 
without actually solving the Einstein's equations 
for that region.
Recall that in the Newtonian theory, the conservation of total 
momentum
immediately follows from Newton's third law,
\begin{equation}
\vec{F}_{\rm total}
={d \over d t}( \sum_{i}\vec{p_{i}} )=0,  \label{newton}
\end{equation}
which is no more than the consistency condition implementing 
Newton's second law. In general relativity, the consistency conditions 
for evolution are already incorporated 
into the Einstein's equations through the constraint equations,
from which global conservation equations were 
found. In this note we will show  that, from the 
Einstein's equations in the (1+1)-dimensional description, 
one can find conservation equations of a stronger 
form, namely, {\it quasi-local} conservation equations. 
These equations, which are integro-differential equations over 
a compact two-dimensional space, are 
naturally interpreted as quasi-local energy, linear momentum, 
and angular momentum conservation equations\cite{wini}.

Let us consider the following line
element
\begin{equation}
ds^2 = -2dudv - 2hdu^2 +{\rm e}^{\sigma} \rho_{ab}
 \left( dy^a + A_{+}^{\ a}du +A_{-}^{\ a} dv \right) 
\left( dy^b + A_{+}^{\ b}du +A_{-}^{\ b} dv 
\right),    \label{yoon}
\end{equation}
where $+,-$ stands for $u,v$, 
respectively
\cite{yoon93a,yoon01,din-sta78,din-sma80,yoon99a,yoon99b,new-unti62,new-tod80}. 
To understand the geometry of this metric, it is convenient to introduce
the following vector fields,
\begin{eqnarray}
& & 
\hat{\partial}_{+}:=\partial_{+} - A_{+}^{\ a}\partial_{a},\label{pplus}\\
& & 
\hat{\partial}_{-}:=\partial_{-} - A_{-}^{\ a}\partial_{a},\label{mminus}
\end{eqnarray}
where we defined the following short-hand notations 
\begin{equation}
\partial_{+}:={\partial \over \partial u}, \hspace{.2in}
\partial_{-}:={\partial \over \partial v}, \hspace{.2in}
\partial_{a}:={\partial  \over \partial y^{a}} \ (a=2,3).
\end{equation}
The inner products of the vector fields 
$\{ \hat{\partial}_{\pm}, \partial_{a} \}$
are given by 
\begin{eqnarray}
& & 
<\hat{\partial}_{+}, \ \hat{\partial}_{+}> = -2h, \hspace{.2in}
<\hat{\partial}_{+}, \ \hat{\partial}_{-}> = -1,  \hspace{.2in} 
<\hat{\partial}_{-}, \ \hat{\partial}_{-}> =0, \nonumber\\
& & 
<\hat{\partial}_{\pm}, \ \partial_{a}> =0, \hspace{.2in}
<\partial_{a}, \ \partial_{b}>={\rm e}^{\sigma} \rho_{ab}.
\end{eqnarray}
The hypersurface $u={\rm constant}$ is a null hypersurface
generated by the out-going null vector field $\hat{\partial}_{-}$,
which is orthogonal to the vector fields $\{ \partial_{a} \}$.
Notice that  $v$ is the {\it affine} parameter of 
the out-going null vector field. 
The hypersurface $v={\rm constant}$ is generated by the 
vector field $\hat{\partial}_{+}$ whose norm is $-2h$, which can be either
negative, zero, or positive. 
The intersection of two hypersurfaces $u,v={\rm constant}$ defines
a spacelike compact two-surface $N_{2}$, which are 
coordinatized by $y^{a}$. The metric on $N_{2}$ is decomposed into
the area element  ${\rm e}^{\sigma}$ and the conformal two-metric
$\rho_{ab}$, which is normalized to have a unit determinant
\begin{equation}
{\rm det}\ \rho_{ab}=1.                        \label{det}
\end{equation}
In the terminology of the fibre bundles, the base manifold
is the (1+1)-dimensional space-time coordinatized by $(u,v)$, 
and the fibre space is 2-dimension spacelike space $N_{2}$. 
The vector fields $\{ \hat{\partial}_{\pm}\} $, which are orthogonal to
$\{ \partial_{a} \}$, is the horizontal vector field, and 
$\{ \partial_{a} \}$ is tangent to the fibre space $N_{2}$. The fields 
$A_{\pm}^{\ a}$ are the corresponding connections 
valued in the diffeomorphisms of the two-surface 
$N_{2}$\cite{yoon92,yoon93a,yoon01}.

For later uses, we shall write down the future-directed 
in-going null vector field $n$ 
and out-going null vector field $l$, orthogonal to two-surface $N_{2}$ 
at each spacetime point. They are given by
\begin{eqnarray}
& & n:= \hat{\partial}_{+} -  h  \hat{\partial}_{-}, \label{en}\\
& & l:= \hat{\partial}_{-},                   \label{el}
\end{eqnarray}
and are normalized such that 
\begin{equation}
< n, \ l > = -1.
\end{equation}
If we further assume that $A_{-}^{\ a}=0$, then the metric 
(\ref{yoon}) becomes identical to the metric studied 
in \cite{new-unti62}. 
In this note, however, we shall retain the $A_{-}^{\ a}$ field,
since its presence will make the $N_{2}$-diffeomorphism invariant
Yang-Mills type gauge theory aspect of this formalism transparent.
Apart from the $N_{2}$-diffeomorphism invariance, there are other 
residual symmetries that preserve the metric (\ref{yoon}),
which are the reparametrization of $u$, and the transformation that 
shifts of the origin of the affine parameter $v$ at each point 
of $N_{2}$\cite{new-tod80}.

The complete set of the vacuum Einstein's equations
are found to be\cite{yoon99b}
\begin{eqnarray}
(a)\hspace{.5cm}  & & {\rm e}^{\sigma} D_{+}D_{-}\sigma 
+ {\rm e}^{\sigma} D_{-}D_{+}\sigma  
+ 2{\rm e}^{\sigma} (D_{+}\sigma)(D_{-}\sigma)
- 2{\rm e}^{\sigma}(D_{-}h)(D_{-}\sigma)  
- {1\over 2}{\rm e}^{ 2 \sigma}\rho_{a b}
F_{+-}^{\ \ a}F_{+-}^{\ \ b}                 \nonumber\\
& & 
+{\rm e}^{\sigma} R_{2}
- h {\rm e}^{\sigma} \Big\{
(D_{-}\sigma)^{2} 
-{1\over 2}\rho^{a b}\rho^{c d} 
 (D_{-}\rho_{a c})(D_{-}\rho_{b d})\Big\}=0, \label{aa}\\
(b)\hspace{.5cm}  & & 
-{\rm e}^{\sigma} D_{+}^{2}\sigma 
- {1\over 2}{\rm e}^{\sigma}(D_{+}\sigma)^{2}
-{\rm e}^{\sigma}(D_{-}h) (D_{+}\sigma)
+{\rm e}^{\sigma}(D_{+}h)(D_{-}\sigma) 
+2h {\rm e}^{\sigma}(D_{-}h)(D_{-}\sigma)  \nonumber\\
& & 
+{\rm e}^{\sigma}F_{+-}^{\ \ a}\partial_{a}h
-{1\over 4}{\rm e}^{\sigma}\rho^{a b}\rho^{c d} 
 (D_{+}\rho_{a c})(D_{+}\rho_{b d})
+\partial_{a}\Big( \rho^{a b}\partial_{b}h \Big) \nonumber\\
& & 
+h\Big\{ - {\rm e}^{\sigma} (D_{+}\sigma) 
  (D_{-}\sigma)
+{1\over 2}{\rm e}^{\sigma}\rho^{a b}\rho^{c d} (D_{+}\rho_{a c})
    (D_{-}\rho_{b d})
+{1\over 2}{\rm e}^{2\sigma}\rho_{a b}F_{+-}^{\ \ a}F_{+-}^{\ \ b}
-{\rm e}^{\sigma}R_{2} \Big\} \nonumber\\
& & 
+h^{2}{\rm e}^{\sigma}\Big\{
(D_{-}\sigma)^{2}
-{1\over 2}\rho^{a b}\rho^{c d}
(D_{-}\rho_{a c}) (D_{-}\rho_{b d})\Big\}=0,  \label{bb}\\
(c)\hspace{.5cm}  & & 
2{\rm e}^{\sigma}(D_{-}^{2}\sigma) +
{\rm e}^{\sigma} (D_{-}\sigma)^{2}  
    + {1\over 2}{\rm e}^{\sigma}\rho^{a b}\rho^{c d} (D_{-}\rho_{a c})
    (D_{-}\rho_{b d})=0,                 \label{cc}\\
(d) \hspace{.5cm}  & &
D_{-}\Big( {\rm e}^{2\sigma}
\rho_{a b}F_{+-}^{\ \ b}\Big)
- {\rm e}^{\sigma}\partial_{a}(D_{-}\sigma) 
- {1\over 2}{\rm e}^{\sigma}\rho^{b c}\rho^{d e}
    (D_{-}\rho_{b d})(\partial_{a}\rho_{c e}) 
+ \partial_{b} \Big(
{\rm e}^{ \sigma}\rho^{b c}D_{-}\rho_{a c} \Big)=0,\label{dd}\\
(e) \hspace{.5cm}  & & 
-D_{+}\Big( {\rm e}^{2\sigma}
\rho_{a b}F_{+-}^{\ \ b}\Big) 
-{\rm e}^{\sigma}\partial_{a} (D_{+}\sigma )
   -{1\over 2}{\rm e}^{\sigma}\rho^{b c}\rho^{d e}
   (D_{+}\rho_{b d})(\partial_{a}\rho_{c e})  
+\partial_{b}\Big(  
{\rm e}^{ \sigma}\rho^{b c}D_{+}\rho_{a c} \Big) \nonumber\\
& & 
+2h{\rm e}^{\sigma}\partial_{a}(D_{-}\sigma)   
+h{\rm e}^{\sigma}\rho^{b c}\rho^{d e}
   (D_{-}\rho_{b d})(\partial_{a}\rho_{c e}) 
+2{\rm e}^{\sigma}\partial_{a}(D_{-}h)    
-2\partial_{b}\Big(h  
{\rm e}^{ \sigma}\rho^{b c}D_{-}\rho_{a c} \Big)
=0,                              \label{ee}\\
(f)\hspace{.5cm}  & & 
-2 {\rm e}^{ \sigma}D_{-}^{2} h 
-2{\rm e}^{ \sigma} (D_{-}h)(D_{-}\sigma) 
+ {\rm e}^{ \sigma}D_{+}D_{-}\sigma 
+ {\rm e}^{ \sigma}D_{-}D_{+}\sigma        
+ {\rm e}^{ \sigma} (D_{+}\sigma)(D_{-}\sigma) \nonumber\\  
& & 
+ {1\over 2}{\rm e}^{ \sigma}\rho^{a b}\rho^{c d} 
  (D_{+}\rho_{a c})(D_{-}\rho_{b d})   
+ {\rm e}^{2 \sigma}\rho_{a b}
    F_{+-}^{\ \ a}F_{+-}^{\ \ b}
 -2h{\rm e}^{ \sigma} \Big\{
   D_{-}^{2} \sigma +{1\over 2}(D_{-}\sigma)^{2} \nonumber\\
& & 
+{1\over 4}\rho^{a b}\rho^{c d}
   (D_{-}\rho_{a c})(D_{-}\rho_{b d})\Big\}=0,  \label{fff}\\
(g)\hspace{.5cm}  & & 
h\Big\{ {\rm e}^{\sigma} D_{-}^{2} \rho_{ab} 
- {\rm e}^{\sigma}\rho^{c d}(D_{-}\rho_{a c})(D_{-}\rho_{b d}) 
+{\rm e}^{\sigma}(D_{-}\rho_{a b})(D_{-}\sigma) \Big\}  \nonumber\\
& & 
-{1\over 2}{\rm e}^{\sigma} \Big( 
D_{+}D_{-}\rho_{a b} + D_{-}D_{+}\rho_{a b} \Big) 
+{1\over 2}{\rm e}^{\sigma} \rho^{c d}\Big\{ 
(D_{-}\rho_{a c})(D_{+}\rho_{b d}) 
+(D_{-}\rho_{b c})(D_{+}\rho_{a d}) \Big\}  \nonumber\\
& & 
-{1\over 2}{\rm e}^{\sigma}\Big\{ 
(D_{-}\rho_{a b})(D_{+}\sigma)
+(D_{+}\rho_{a b})(D_{-}\sigma)  \Big\}  \nonumber\\
& & 
 +{\rm e}^{\sigma}(D_{-}\rho_{a b})(D_{-}h)                
+{1\over 2}{\rm e}^{2 \sigma}\rho_{a c}\rho_{b d}
 F_{+-}^{\ \ c}F_{+-}^{\ \ d}     
-{1\over 4}{\rm e}^{2 \sigma}\rho_{a b}
 \rho_{c d}F_{+-}^{\ \ c}F_{+-}^{\ \ d}=0. \label{ggg}
\end{eqnarray}
Here $R_{2}$ is the scalar curvature of $N_{2}$, and 
we defined the diff$N_{2}$-covariant derivatives as follows,
\begin{eqnarray}
& &F_{+-}^{\ \ a}:=\partial_{+} A_{-} ^ { \ a}-\partial_{-}
  A_{+} ^ { \ a} - [A_{+}, A_{-}]_{\rm L}^{a},  \label{field}\\
& &D_{\pm}\sigma := \partial_{\pm}\sigma
-[A_{\pm}, \sigma]_{\rm L},         \label{get}\\
& &D_{\pm}h:= \partial_{\pm}h - [A_{\pm}, h]_{\rm L},   \label{eichid}\\
& &D_{\pm}\rho_{a b}:=\partial_{\pm}\rho_{a b}
   - [A_{\pm}, \rho]_{{\rm L}a b}.    \label{rhod}
\end{eqnarray}
The bracket $[A_{\pm}, f]_{{\rm L} a b\cdots }$ is the Lie 
derivative  
of $f_{a b\cdots }$ along the vector field 
$A_{\pm}:=A_{\pm}^{\ a}\partial_{a}$, defined as
\begin{equation}
[A_{\pm}, f]_{{\rm L}a b\cdots}
:=A_{\pm}^{\ c}\partial_{c}f_{ab\cdots}
+f_{cb\cdots}\partial_{a}A_{\pm}^{\ c}
+f_{ac\cdots}\partial_{b}A_{\pm}^{\ c}
-w (\partial_{c}A_{\pm}^{\ c})f_{ab\cdots},
\end{equation}
where $w$ is the weight of the tensor density $f_{ab \cdots}$.
One can also compute the scalar curvature $R$ of the metric 
(\ref{yoon})
and integrate it over spacetime. It is given by
\begin{eqnarray}
I_{0} 
&=&   \int \! \! du \, dv \, d^{2}y \, {\rm e}^{\sigma} R
                               \nonumber\\
&=& \int \! \! du \, dv \, d^{2}y \, L_{0} 
+ {\rm surface}  \ {\rm integrals},       \label{bareact}
\end{eqnarray}
where the ``Lagrangian'' function 
$L_{0}$ is given by\cite{yoon93a}
\begin{eqnarray}
& & L_{0} = -{1\over 2}{\rm e}^{2 \sigma}\rho_{a b}
  F_{+-}^{\ \ a}F_{+-}^{\ \ b}
  +{\rm e}^{\sigma} (D_{+}\sigma) (D_{-}\sigma)
  -{1\over 2}{\rm e}^{\sigma}\rho^{a b}\rho^{c d}
 (D_{+}\rho_{a c})(D_{-}\rho_{b d})
 -{\rm e}^{\sigma} R_2            \nonumber\\
& & -2{\rm e}^{\sigma}(D_{-}h)(D_{-}\sigma) 
- h {\rm e}^{\sigma}(D_{-}\sigma)^2
+{1\over 2}h  {\rm e}^{\sigma}\rho^{a b}\rho^{c d}
 (D_{-}\rho_{a c})(D_{-}\rho_{b d}).    \label{barelag}
\end{eqnarray}

One can easily recognize that this ``Lagrangian'' function  $L_{0}$ 
is in a form of a (1+1)-dimensional field theory Lagrangian.
In geometrical terms the function $L_{0}$ describes
how the (1+1)-dimensionsal spacetime and 2-dimensional fibre
space are imbedded into an enveloping (3+1)-dimensional spacetime.
Each term in (\ref{barelag}) is 
manifestly diff$N_{2}$-invariant, and the $y^{a}$-dependence 
of each term is completely ``hidden'' in the Lie derivatives.
In this sense we may regard the fibre space $N_{2}$ as 
a kind of ``internal'' space as in a Yang-Mills theory,
with the infinite dimensional group of diffeomorphism of $N_{2}$ as 
the Yang-Mills gauge symmetry. 
Thus, the above function $L_{0}$ is describable as 
a (1+1)-dimensional Yang-Mills type gauge theory interacting with 
(1+1)-dimensional scalar fields and non-linear sigma fields 
of generic types.

\end{section}

\begin{section}{A Set of Quasi-local Conservation Equations}

Notice that the four equations (\ref{aa}), (\ref{bb}) and (\ref{ee}) 
are partial differential equations that are {\it first-order} 
in $D_{-}$ derivatives. Therefore it is of particular interest to study 
these four equations, since they are close analogues to
the Einstein's constraint equations in the usual (3+1) formalism. 
Thus, in this formalism, 
the {\it natural} vector field that defines the evolution
is $D_{-}$.
Then the momenta 
\begin{math}
\pi_{I}=\{ \pi_{h},\pi_{\sigma}, \pi_{a}, \pi^{a b} \} 
\end{math} \
conjugate to the configuration variables 
\begin{math}
q^{I}=\{ h, \sigma,   A_{+} ^ { \ a}, \rho_{a b} \}
\end{math} \
are {\it defined} as 
\begin{equation}
\pi_{I}:={\partial L_{0}\over \partial (D_{-}{q}^{I}) }.
\label{momenta}
\end{equation}
They are found to be
\begin{eqnarray}
& &\pi_{h}=-2 {\rm e}^{\sigma}(D_{-}\sigma),   \label{pih}\\
& &\pi_{\sigma} = -2 {\rm e}^{\sigma} (D_{-}h)
       -2h {\rm e}^{\sigma} (D_{-}\sigma)
     + {\rm e}^{\sigma} (D_{+}\sigma),   \label{pisigma}  \\
& &\pi_{a}={\rm e}^{2 \sigma} \rho_{a b}F_{+-}^{\ \ b}, \label{pia} \\
& &\pi^{a b}= 
h{\rm e}^{\sigma} \rho^{a c}\rho^{b d}(D_{-}\rho_{c d})
-{1\over 2}{\rm e}^{\sigma} \rho^{a c}\rho^{b d}
(D_{+}\rho_{c d}).                            \label{definition}
\end{eqnarray}
Notice that  $\pi^{a b}$ is traceless
\begin{equation}
\pi^{a}_{\ a}=0,
\end{equation}
due to the identities that are direct consequences of 
the condition
(\ref{det}), 
\begin{equation}
\rho^{ab}D_{\pm}\rho_{ab}=0.                   \label{module} 
\end{equation} 
The ``Hamiltonian'' function $H_{0}$ {\it defined} as 
\begin{equation}
H_{0}:=\pi_{I}D_{-}{q}^{I} - L_{0}
\end{equation}
is found to be
\begin{equation}
H_{0}= H + {\rm total}\ {\rm  divergences}, \label{ok}
\end{equation}                             
where $H$ is given by
\begin{eqnarray}
& & H  =  -{1\over 2}{\rm e}^{-\sigma}\pi_{\sigma}\pi_{h} 
+ {1\over 4}h{\rm e}^{-\sigma}\pi_{h}^{2} 
-{1\over 2}{\rm e}^{-2\sigma}\rho^{a b}\pi_{a}\pi_{b} 
+{1\over 2h}{\rm e}^{-\sigma}
\rho_{a c}\rho_{b d}\pi^{a b}\pi^{c d}    \nonumber\\
& &
+{1\over 2}\pi_{h}(D_{+}\sigma)  
+{1\over 2h}\pi^{a b}(D_{+}\rho_{a b}) 
+{1\over 8h}{\rm e}^{\sigma}\rho^{a b} \rho^{c d}
(D_{+}\rho_{a c}) (D_{+}\rho_{b d}) 
+{\rm e}^{\sigma}R_{2}.        \label{tilde}
\end{eqnarray}
In terms of these canonical variables $\{\pi_{I},{q}^{I}\}$,
the first-order equations (\ref{aa}), (\ref{bb}), and 
(\ref{ee}) can be written as, after a little algebra, 
\begin{eqnarray}
({\rm i}) \hspace{.5cm} & & 
\pi^{a b}D_{+}\rho_{a b}   + \pi_{\sigma}D_{+}\sigma 
-h D_{+} \pi_{h}    
-\partial_{+}\Big( 
     h\, \pi_{h} + 2 {\rm e}^{\sigma}D_{+}\sigma \Big) \nonumber\\
& &
+\partial_{a}\Big( 
    h\, \pi_{h}A_{+}^{ \ a}    
+ 2A_{+}^{ \ a}{\rm e}^{\sigma}D_{+}\sigma 
 + 2h {\rm e}^{-\sigma}\rho^{a b}\pi_{b}  
 +2\rho^{a b}\partial_{b}h \Big) =0,         \label{qenergy}\\
({\rm ii}) \hspace{.5cm} & & 
H - \partial_{+}\pi_{h} 
+ \partial_{a} \Big( 
A_{+}^{\ a}\pi_{h} 
+ {\rm e}^{-\sigma}\rho^{a b} \pi_{b} \Big)=0,  \label{qmomentum}\\
({\rm iii}) \hspace{.5cm} & & 
\partial_{+}\pi_{a} 
-\partial_{b}(A_{+}^{\ b}\pi_{a})
-\pi_{b}\partial_{a}A_{+}^{\ b}
-\pi_{\sigma}\partial_{a}\sigma 
+ \partial_{a}\pi_{\sigma} - \pi_{h}\partial_{a}h
     - \pi^{b c}\partial_{a}\rho_{b c}   \nonumber\\
& &      + \partial_{b}( \pi^{b c}\rho_{a c})    
     + \partial_{c}( \pi^{b c}\rho_{a b})   
     - \partial_{a}( \pi^{b c}\rho_{b c})=0.  \label{qangular}
\end{eqnarray}
These are the four first-order equations in the gauge (\ref{yoon}),
and it is these equations that we are concerned with in this note.
Notice that the equations (\ref{qenergy}) and (\ref{qmomentum}) 
are {\it divergence}-type equations.
If we contract the equations (\ref{qangular}) 
by an arbitrary function $\xi^{a}$ of $\{ v, y^{b} \}$ such that
\begin{equation}
\partial_{+} \xi^{a}=0,                 \label{xicon}
\end{equation}
then the resulting equation is also a divergence-type equation, 
\begin{equation}
\pi^{a b}{\pounds_{\xi}} \rho_{a b}
+\pi_{\sigma}{\pounds_{\xi}} \sigma
+\pi_{h}{\pounds_{\xi}} h
+\pi_{a}{\pounds_{\xi}} A_{+}^{\ a} 
-\partial_{+}( \xi^{a}\pi_{a} )   
+\partial_{a}\Big(
-\xi^{a}  \pi_{\sigma} + 2 \pi^{a b} \xi^{c} \rho_{b c}
+A_{+}^{\ a} \xi^{b} \pi_{b} \Big)
=0,                                   \label{qangular2}
\end{equation}
where $\pounds_{\xi}$ is the Lie derivative 
along the vector field $\xi:=\xi^{a}\partial_{a}$.

The integrals of these equations over a compact two-surface 
$N_{2}$ become, after the normalization by $1/ 16\pi$, 
\begin{eqnarray}
& & {\partial \over \partial u} U(u,v)
={1\over 16\pi}  \oint \! d^{2}y \, \Big(
\pi^{a b}D_{+}\rho_{a b} +\pi_{\sigma}D_{+}\sigma  
-h D_{+} \pi_{h}
   \Big),              \label{enflux} \\
& &  {\partial \over \partial u} P(u,v)
={1 \over 16\pi} \oint \! d^{2}y \, H,  \label{momflux}  \\
& & {\partial \over \partial u} L(u,v;\xi)
={1 \over 16\pi}\oint \! d^{2}y \, \Big( 
\pi^{a b}{\pounds}_{\xi} \rho_{a b}  
+\pi_{\sigma}{\pounds}_{\xi} \sigma 
-h {\pounds}_{\xi} \pi_{h} 
-A_{+}^{\ a}{\pounds}_{\xi}\pi_{a} \Big) 
\hspace{.25in} (\partial_{+} \xi^{a}=0),   \label{angflux} 
\end{eqnarray}
where in the last integral we used the fact that 
\begin{equation}
\oint \! d^{2}y \,  {\pounds_{\xi}} f 
=\oint \! d^{2}y \,  \partial_{a} (\xi^{a}f) 
=0
\end{equation}
for a scalar density $f$ with the weight $-1$. 
Here $U(u,v)$, $P(u,v)$, and $L(u,v;\xi)$ are invariant two-surface 
integrals defined as
\begin{eqnarray}
& & U(u,v):= {1 \over 16\pi}\oint d^{2}y \,   \Big( 
h\, \pi_{h} + 2 {\rm e}^{\sigma} D_{+}\sigma \Big) 
+ \bar{U},               \label{enint}\\
& &
P(u,v):={1 \over 16\pi} \oint \! d^{2}y \, ( \pi_{h} )
+ \bar{P},                \label{momint}\\
& &
L(u,v;\xi):={1 \over 16\pi} \oint \! d^{2}y \, 
(\xi^{a}\pi_{a}) + \bar{L}    
\hspace{.25in} (\partial_{+} \xi^{a}=0),   \label{angint}
\end{eqnarray}
where $\bar{U}$, $\bar{P}$, and $\bar{L}$ are undetermined 
subtraction terms. Notice that these subraction 
terms must be $u$-independent,
\begin{equation}
{\partial \bar{U}\over \partial u} 
={\partial \bar{P}\over \partial u}
={\partial \bar{L}\over \partial u} =0,    \label{subtract}
\end{equation}
in order to satisfy the equations (\ref{enflux}), (\ref{momflux}), 
and (\ref{angflux}), respectively. 
In general the subtraction terms  are
not unique, and the ``right'' subtraction term may not even exist 
at all in a generic situation. One natural criterion for the 
``right'' choice of subtraction term would be 
that it must be chosen such that the 
quasi-local physical quantities reproduce ``standard'' values 
in the well-known limiting cases.

One can write the r.h.s. of the equation (\ref{enflux}) 
in a more symmetric and suggestive form as follows.
To do this, let us contract the equation (\ref{qangular}) with $A_{+}^{\ a}$
and integrate over $N_{2}$ 
to obtain the following equation
\begin{equation}
\oint \! d^{2}y \, \Big( 
A_{+}^{\ a} \partial_{+}\pi_{a} \Big)  
=\oint \! d^{2}y \, \Big( 
\pi^{a b}{\pounds}_{A_{+}} \rho_{a b}  
+\pi_{\sigma}{\pounds}_{A_{+}} \sigma 
-h {\pounds}_{A_{+}} \pi_{h} \Big).       \label{xxx} 
\end{equation}
If we use the definition of diff$N_{2}$-covariant 
derivatives $D_{\pm}$ and 
the equation (\ref{xxx}), then the equation (\ref{enflux}) 
can be written as
\begin{equation}
{\partial \over \partial u} U(u,v)
={1\over 16\pi}  \oint \! d^{2}y \, \Big(
\pi^{a b}  \partial_{+} \rho_{a b} 
+\pi_{\sigma}\partial_{+}\sigma  - h \partial_{+} \pi_{h}
-A_{+}^{\ a} \partial_{+}\pi_{a} \Big),   \label{enflux1}
\end{equation}
where the integrand on the r.h.s. assumes the canonical form of 
energy-flux, which is typically given by 
\begin{equation}
T_{0+} \sim \sum_{i}\pi_{i}\partial_{+}\phi^{i}, 
\end{equation}
where $\phi{^i}$ is a generic field and $\pi_{i}$ is its conjugate
momentum. 
Notice that the r.h.s. of the conservation equations (\ref{angflux}) 
and (\ref{enflux1}) match exactly, if we interchange the 
derivatives in the integrands
\begin{equation}
{\pounds}_{\xi}  \longleftrightarrow \partial_{+}. \label{match}
\end{equation}
In a region of a spacetime where
\begin{math}
\partial / \partial u
\end{math} \ 
is timelike, these quasi-local equations becomes
quasi-local conservation equations, which 
relate the instantaneous 
rates of changes of two-surface integrals at a given $u$-time
to the associated net flux integrals. Let us remark that,
unlike the Tamburino-Winicour's quasi-local conservation 
equations\cite{wini} which are ``weak'' conservation equations 
since  the Ricci flat conditions (i.e. the full vacuum Einstein's 
equations) were assumed in their derivation, 
our quasi-local conservation equations are ``strong'' 
conservation equations since only the four first-order 
equations were used in the derivation.

It is interesting to notice that 
we can obtain yet another quasi-local conservation equation.
This is simply achieved by writing the equation (\ref{xxx}) as 
\begin{equation}
{\partial \over \partial u} 
\oint \! d^{2}y \, ( 
A_{+}^{\ a} \pi_{a} )
=
\oint \! d^{2}y \, \Big( 
\pi^{a b}{\pounds}_{A_{+}} \rho_{a b}  
+\pi_{\sigma}{\pounds}_{A_{+}} \sigma 
-h {\pounds}_{A_{+}} \pi_{h} 
+ \pi_{a}\partial_{+}A_{+}^{\ a} \Big),   \label{xxxa} 
\end{equation}
which relates the instantaneous $u$-derivative of the two-surface
integral on the l.h.s. to the net flux integral on the right. 
However, the r.h.s. of this equation is not quite ``canonical'' due to 
the last term. If we restrict the field $A_{+}^{\ a}$  such that it 
satisfies the $u$-independent condition 
\begin{equation}
\partial_{+}A_{+}^{\ a}=0,                 \label{aplus}
\end{equation}
which is essentially the same condition (\ref{xicon})
that $\xi^{a}$ satisfies, then the last term in the r.h.s. of 
the equation (\ref{xxxa}) drops out, and we obtain the following equation
\begin{equation}
{\partial \over \partial u} J(u,v) 
={1\over 16\pi} 
\oint \! d^{2}y \, \Big( 
\pi^{a b}{\pounds}_{A_{+}} \rho_{a b}  
+\pi_{\sigma}{\pounds}_{A_{+}} \sigma 
-h {\pounds}_{A_{+}} \pi_{h} \Big).          \label{xxxb} 
\end{equation}
Here $J(u,v)$ is defined as 
\begin{equation}
J(u,v):= {1\over 16\pi} \oint \! d^{2}y \, ( 
A_{+}^{\ a} \pi_{a} ) + \bar{J}
\hspace{.25in}
(\partial_{+}A_{+}^{\ a}=0),           \label{xxxint}
\end{equation}
where $\bar{J}$ is an undetermined subtraction term.
The r.h.s. of the equation (\ref{xxxb}) now represents a flux of 
the canonical form
\begin{equation}
\sum_{i}\pi_{i}{\pounds}_{A_{+}}\phi^{i},
\end{equation}
just as the r.h.s. of the equations (\ref{angflux}) and (\ref{enflux1})
do.

\end{section}

\begin{section}{Geometrical interpretations}

Remarkably, the two-surface integrals (\ref{enint}), (\ref{momint}), 
(\ref{angint}), and (\ref{xxxint}), which were derived 
using the metric (\ref{yoon}), can be expressed geometrically, 
in terms of the area of the two-surface and a pair of in-going and 
out-going null vector fields orthogonal to that surface. 
In order to show this, we need to invoke the definitions of 
in-going and out-going null vector fields $\{ n, \ l\}$
defined in the section \ref{intro}.

\begin{subsection}{Quasi-local energy}

Let us first observe that the integral in (\ref{enint}) can be written as 
the Lie derivative of 
the scalar density ${\rm e}^{\sigma}$ along the in-going null 
vector field $n$,
\begin{eqnarray} 
\oint d^{2}y \, \Big( h\, \pi_{h} + 2 {\rm e}^{\sigma} D_{+}\sigma \Big)
& = &  2 \oint d^{2}y \,  
{\rm e}^{\sigma} \Big (D_{+}\sigma - h D_{-}\sigma \Big) \nonumber\\
& = &  2 \oint d^{2}y \, {\pounds}_{n}{\rm e}^{\sigma}.  \label{triv}
\end{eqnarray}
One finds that 
\begin{equation}
\oint d^{2}y \, {\pounds}_{n}{\rm e}^\sigma
= {\pounds}_{n} {\cal A},                          \label{inter}
\end{equation}
where ${\cal A}$ is the area of $N_{2}$, 
\begin{equation}
{\cal A}=\oint d^{2}y \,{\rm e}^\sigma.
\end{equation}
The identity (\ref{inter}) follows trivially from the observation that 
the null vector field $n$ is out of 
(in fact, orthogonal to) the two-surface, which means that the order of 
the integration over $d^{2}y$ and the Lie derivative  ${\pounds}_{n}$ 
in (\ref{triv}) is interchangeable. 
Thus we have 
\begin{equation}
{1 \over 16\pi}\oint d^{2}y \,   \Big( 
h\, \pi_{h} + 2 {\rm e}^{\sigma} D_{+}\sigma \Big)
={1 \over 8\pi} {\pounds}_{n}{\cal A}. 
\end{equation}
For a reference term $\bar{U}$, let us choose 
\begin{equation}
\bar{U}:=-{1 \over 8\pi} {\pounds}_{\bar{n}}{\cal A},
\end{equation}
where $\bar{n}$ is a future-directed in-going null vector field
of a background reference spacetime $d\bar{s}^2$
into which the two-surface $N_{2}$ (with the same metric 
${\rm e}^{\sigma} \rho_{ab}$) is embedded, 
\begin{equation}
d\bar{s}^2
= -2dudv - 2\bar{h}du^2 +{\rm e}^{\sigma} \rho_{ab}
 \left( dy^a + \bar{A}_{+}^{\ a}du +\bar{A}_{-}^{\ a} dv \right) 
\left( dy^b + \bar{A}_{+}^{\ b}du +\bar{A}_{-}^{\ b} dv 
\right).                         \label{yoonref}
\end{equation}
Notice that $\bar{n}$, which is given by  
\begin{equation}
\bar{n}:=\Big( {\partial \over \partial u} 
- \bar{A}_{+}^{\ a}{\partial \over \partial y^{a}} \Big)
-\bar{h}  \Big( 
{\partial \over \partial v} 
- \bar{A}_{-}^{\ a}{\partial \over \partial y^{a}}\Big),
                             \label{enbar}
\end{equation}
is a function of $\bar{h}$ and $\bar{A}_{\pm}^{\ a}$ only,
the embedding degrees of freedom of the two-surface.
Thus, the quasi-local energy of a given two-surface
$N_{2}$ is defined relative to some fixed background reference 
spacetime, and is zero when the two-surface under consideration 
is embedded into the fixed background reference spacetime, 
i.e. when
\begin{equation}
n=\bar{n}.
\end{equation}
Therefore, the quantity $U(u,v)$, which will be interpreted as 
the quasi-local energy, becomes
\begin{equation}
U(u,v):= {1 \over 8\pi} {\pounds}_{n}{\cal A}
-{1 \over 8\pi} {\pounds}_{\bar{n}}{\cal A}.    \label{eninta}
\end{equation}
It is given by the rate of change of the area of a given 
two-surface along the future-directed {\it in}-going null vector field
$n$, relative to some background null vector field $\bar{n}$. 
Notice that this definition is entirely geometrical, referring to 
the area of a given two-surface and the orthogonal null vector fields 
$n$ (and $\bar{n}$) only.

\end{subsection}

\begin{subsection}{Quasi-local linear momentum}

The two-surface integral (\ref{momint}) can be also written 
geometrically in a similar way. It becomes 
\begin{equation}
{1 \over 16\pi} \oint \! d^{2}y \, ( \pi_{h} )
=-{1 \over 8\pi} \oint \! d^{2}y \, {\rm e}^{\sigma} D_{-}\sigma
=-{1 \over 8\pi} \oint \! d^{2}y \, {\rm e}^{\sigma} 
 {\pounds}_{l} \sigma
=-{1 \over 8\pi}{\pounds}_{l}{\cal A}.     \label{mominta}
\end{equation}
Therefore, if we choose the reference term  $\bar{P}$
as 
\begin{equation}
\bar{P}:={1 \over 8\pi}{\pounds}_{\bar{l}}{\cal A},
\end{equation}
where 
$\bar{l}$ is a future-directed out-going null 
vector field of a background reference spacetime, 
then $P$ becomes,
\begin{equation}
P(u,v)=-{1 \over 8\pi}{\pounds}_{l}{\cal A} 
+{1 \over 8\pi}{\pounds}_{\bar{l}}{\cal A}.  \label{momintb}
\end{equation}
Thus, the quasi-local integral $P(u,v)$, which is to be
interpreted as the quasi-local linear momentum, 
is given by the rate of change of the area of a given 
two-surface along the future-directed {\it out}-going null vector 
field $l$ relative to $\bar{l}$.

\end{subsection}

\begin{subsection}{Quasi-local angular momentum}

Let us also write down the two-surface integral (\ref{angint}) in a 
geometrical way. Notice that the Lie bracket of the two null vector 
fields $\{n,l \}$ is given by 
\begin{equation}
[n,\ l ]_{\rm L}
=-F_{+-}^{ \ \ a}\partial_{a} + (D_{-}h) l.
\end{equation}
Thus, (\ref{angint})  becomes
\begin{eqnarray}
{1 \over 16\pi} \oint \! d^{2}y \, 
(\xi^{a}\pi_{a})
&=&{1 \over 16\pi} \oint \! d^{2}y \, 
{\rm e}^{ \sigma} \xi_{a} F_{+-}^{\ \ a}   \nonumber\\
&=&-{1 \over 16\pi} \oint \! d^{2}y \, 
{\rm e}^{ \sigma} \xi_{a}  [n, \ l ]_{\rm L}^{a},  \label{twist}
\end{eqnarray}                               
where
\begin{equation}
\xi_{a}:= \phi_{a b}\xi^{b} 
= {\rm e}^{ \sigma} \rho_{a b} \xi^{b}.
\end{equation}
This shows that (\ref{twist}) is an invariant two-surface 
integral of the Lie-bracket $[n,\ l]_{\rm L}$ projected to
the spacelike vector field $\xi:=\xi^{a}\partial_{a}$. 
If we choose the reference term $\bar{L}$ as
\begin{equation}
\bar{L}:={1 \over 16\pi} \oint \! d^{2}y \, 
{\rm e}^{ \sigma} \xi_{a}  [\bar{n}, \ \bar{l} ]_{\rm L}^{a},
                                      \label{elref}
\end{equation}
then (\ref{angint}) becomes
\begin{equation}
L(u,v;\xi)
=-{1 \over 16\pi} \oint \! d^{2}y \, 
{\rm e}^{ \sigma} \xi_{a}  [n, \ l ]_{\rm L}^{a} 
+{1 \over 16\pi} \oint \! d^{2}y \, 
{\rm e}^{ \sigma} \xi_{a}  [\bar{n}, \ \bar{l} ]_{\rm L}^{a}
\hspace{.25in} 
(\partial_{+}\xi^{a}=0).             \label{goodef}
\end{equation}
It must be stressed $\xi^{a}$ is an arbitrary function of $\{v, y^{b}\}$. 
In particular, it need not satisfy any Killing's equations
associated with isometries of a spacetime, or 
of a two-surface within a given spacetime. The quantity $L(u,v;\xi)$,
a linear functional of $\xi^{a}$, can be interpreted as the quasi-local 
angular momentum of a two-surface associated with an arbitrary 
function $\xi^{a}$ (of $y^{b})$, as we will see later.

\end{subsection}

\begin{subsection}{Quasi-local Carter's constant}

Likewise, the fourth integral (\ref{xxxint}) can be written as
\begin{equation}
J(u,v)
=-{1 \over 16\pi} \oint \! d^{2}y \, 
{\rm e}^{ 2\sigma} \rho_{a b} A_{+}^{\ a}  [n, \ l ]_{\rm L}^{b} 
+{1 \over 16\pi} \oint \! d^{2}y \, 
{\rm e}^{ 2\sigma}  \rho_{a b} A_{+}^{\ a} 
[\bar{n}, \ \bar{l} ]_{\rm L}^{b}  
\hspace{.25in}
(\partial_{+}A_{+}^{\ a}=0),             \label{goodefa}
\end{equation}
which may be interpreted as a quasi-local, finite, analog of 
the Carter's ``fourth'' constant, as we shall see in the next section.

\end{subsection}

\end{section}

\begin{section}{Asymptotically flat limits}{\label{k2}}

The equations (\ref{enflux}), (\ref{momflux}), and 
(\ref{angflux}) turn out to be
quasi-local energy, linear momentum, and angular momentum 
conservation equations, respectively\cite{yoon01},
and the equation (\ref{xxxb}) is interpreted as 
a quasi-local conservation equation of the generalized Carter's 
constant\cite{carter68,hugh-pen-som-walk72,carter79}. 
In this section we shall evaluate the two-surface integrals
(\ref{eninta}), (\ref{momintb}), (\ref{goodef}) and 
the associated flux integrals in the limiting asymptotically flat
region where $N_{2}=S_{2}$, 
and show that they all reduce to the well-known Bondi energy, 
linear momentum, angular momentum, and the corresponding 
flux integrals defined at the null infinity.
The asymptotic form of the ``fourth'' integral
(\ref{goodefa}) at the null infinity will be also computed, 
and it will be shown that it is proportional
to the total angular momentum squared. 

In the limit where the affine parameter $v$ 
approaches to infinity, 
the asymptotic form of the Kerr metric becomes 
\begin{eqnarray}
ds^2 & & \longrightarrow  
-2 du dv
-\Big( 1-{2m\over v} +\cdots \Big)\, du^2 
+ \Big( {4 m a  {\rm sin}^{2}\vartheta \over v}
-{4 m a^{3}  {\rm sin}^{2}\vartheta   {\rm cos}^{2}\vartheta \over v^{3}}
 + \cdots \Big) du d \varphi           \nonumber\\
& & 
+v^{2} 
\Big( 1 + {a^2  {\rm cos}^{2}\vartheta \over v^{2}} + \cdots \Big)
d\vartheta^{2} 
+v^{2}  {\rm sin}^{2}\vartheta 
\Big( 1 + {a^2\over v^{2}} + \cdots \Big) d\varphi^{2}  \nonumber\\
& & 
+  {\rm sin}^{2}\vartheta \Big( 
{4 m a^3  \over v^{3}} +{8 m^{2} a^3  \over v^{4}}+ \cdots \Big) 
dv d\varphi
-\Big( {a^2   {\rm sin}^{2}\vartheta \over v^{2}} + \cdots \Big)
dv^{2},                                         \label{sola}
\end{eqnarray}
where
\begin{math}
\partial / \partial u
\end{math} \ 
is asymptotic to the timelike Killing vector field at infinity.
The asymptotic fall-off rates of the metric coefficients 
can be read off from the above 
metric\cite{BMS62,prior77,wald84,penrose86,chrusciel02}, 
\begin{eqnarray}
& &{\rm e}^{\sigma}=v^{2}   ({\rm sin}\vartheta) \Big\{ 
1 +O({1 \over v^{2}})  \Big\},           \label{fallsig}\\
& &\rho_{\vartheta \vartheta}
=\Big( {1\over {\rm sin}\vartheta} \Big) \Big\{ 
1 + { C(u,\vartheta,\varphi)\over v}
+ O({1 \over v^{2}}) \Big\},       \label{fallrho1} \\  
& &\rho_{\varphi \varphi}=({\rm sin}\vartheta)  \Big\{ 
    1 - { C(u,\vartheta,\varphi) \over v} 
+ O({1 \over v^{2}}) \Big\},    \label{fallrho2} \\
& &\rho_{\vartheta \varphi}
=O({1\over v^{2}}),   \label{fallrho3}\\
& &2h=1-{2m \over v} 
 + O({1 \over  v^{2}}),         \label{fallh}\\
& &A_{+}^{\ \varphi}={2ma \over v^{3}} + O({1 \over v^{4}}), \\
& & A_{-}^{\ \varphi}= {2ma^{3} \over v^{5}} 
+ O({1 \over v^{6}}),                      \\
& & A_{\pm}^{\ \vartheta}=  O({1 \over v^{6}}).     \label{falla}
\end{eqnarray}
From these asymptotic behaviors, we can deduce the fall-off rates
of the following derivatives,
\begin{eqnarray}
& & 
\partial_{+}\sigma = O({1\over v^{2}}), \hspace{.2in}
\partial_{-}\sigma = {2\over v}+ O({1\over v^{2}}), \hspace{.2in}
\partial_{+}\rho_{a b} = O({1\over v}), \hspace{.2in}
\partial_{-}\rho_{a b} = O({1\over v^{2}}), \hspace{.2in}
{\pounds}_{\xi} \rho_{a b} = O({1\over v}),  \nonumber\\
& & 
\pi_{h}=-4v{\rm sin}\vartheta + O(1),    \hspace{.2in} 
\pi_{\sigma}=-2v{\rm sin}\vartheta + O(1), \hspace{.2in} 
\pi^{a b} = -{1\over 2}{\rm e}^{\sigma}\rho^{a c}\rho^{b d}
 (\partial_{+}\rho_{c d}) +  O(1), \nonumber\\
& & 
\pi_{\varphi}= 6ma {\rm sin}^{3}\vartheta + O({1\over v}),  \hspace{.2in} 
\pi_{\vartheta}= O({1\over v^{2}}).  
\end{eqnarray}

\begin{subsection}{The Bondi energy-loss relation}

Since the integrand of the r.h.s. of (\ref{enflux1}) assumes the typical
form of energy-flux, we expect that it represents 
the energy-flux carried by gravitational radiation crossing $S_{2}$. 
Then the l.h.s. of (\ref{enflux1}) should be 
the instantaneous rate of change in the gravitational energy 
of the region enclosed by $S_{2}$. 
The energy-flux integral in general does not have a definite sign, 
since it includes the energy-flux carried by the in-coming 
as well as the out-going gravitational radiation. 
But in the asymptotically flat region, 
the energy-flux integral turns out to be negative-definite, 
representing the physical situation that 
there is no in-coming flux coming from the infinity.

Let us now show that the equation (\ref{enflux1}) 
reduces to the Bondi energy-loss formula\cite{new-tod80}
in the asymptotic region of asymptotically flat spacetimes.
To show this, let us first calculate $U(u,v)$ in the limit
$v\longrightarrow \infty$. 
Since the null vector fields $n$ and $l$ asymptotically approach to 
\begin{eqnarray}
& & 
n \longrightarrow {\partial}_{+} 
-  \Big({1\over 2}-{m\over v}\Big) 
  {\partial}_{-},    \nonumber\\
& & 
l \longrightarrow {\partial}_{-},
\end{eqnarray}
the natural background spacetime is
the flat spacetime so that the embedding degrees of freedom are 
given by
\begin{equation}
\bar{A}_{\pm}^{ \ \ a}=0, \hspace{.25in}
2\bar{h}=1.
\end{equation}
That is, the background fields $\bar{n}$ and  $\bar{l}$ become
\begin{equation}
\bar{n}= {\partial}_{+} -  {1\over 2}\partial_{-}, 
\hspace{.2in}
\bar{l}= {\partial}_{-}.                \label{backn}
\end{equation}
Then it follows trivially that 
the total energy at the null infinity coincides with 
the Bondi energy $U_{\rm B}(u)$,
\begin{equation}
\lim_{v \rightarrow \infty}U(u,v) 
:=U_{\rm B}(u) = m,
\end{equation}
where $m$ is the Bondi mass of asymptotically flat spacetimes.
One can further show that the equation (\ref{enflux1}) 
is just the Bondi energy-loss formula, 
\begin{equation} 
{d \over d u} U_{\rm B}(u) 
= -\lim_{v \rightarrow \infty}
{1\over 32\pi}  \oint_{S_{2}} \!\!\! 
d \Omega \, \, v^{2}
  \rho^{a b}\rho^{c d}
(\partial_{+}\rho_{a c})(\partial_{+}\rho_{b d}), \label{lindoo}
\end{equation}
or, equivalently,
\begin{equation} 
{d \over d u} U_{\rm B}(u) 
=-{1\over 16\pi}  \oint_{S_{2}} \!\!\! 
d \Omega \, \, (\partial_{+}C)^{2}
\leq 0,                            \label{lindo}
\end{equation} 
where we used the expressions (\ref{fallrho1}) and (\ref{fallrho2}). 
Notice that the negative-definite energy-flux is a bilinear of 
the traceless {\it current} $j^{a}_{ \ b}$ defined as
\begin{equation}
j^{a}_{ \ b}:=\rho^{a c}\partial_{+}\rho_{ b c} \hspace{.5cm}
(j^{a}_{ \ a}=0), 
\end{equation}
representing the shear degrees of freedom 
of gravitational radiation.

\end{subsection}

\begin{subsection}{The Bondi linear momentum and linear 
momentum-flux}

Let us now evaluate $P(u,v)$ in (\ref{momintb})
and the corresponding quasi-local momentum flux integral 
in the asymptotic region of asymptotically flat spacetimes. 
We find that the total linear momentum $P(u,v)$ becomes zero 
in the asymptotic limit, 
\begin{equation}
\lim_{v\rightarrow \infty} P(u,v)
:=P_{\rm B}(u) =0,                          \label{bondim}
\end{equation}
from which we infer that the total momentum flux 
is zero, 
\begin{equation}
{d \over d u} P_{\rm B}(u)=0.               \label{bondif}
\end{equation}
The result (\ref{bondif}) can be also obtained 
by evaluating each term in the ``Hamiltonian'' function 
$H$ (\ref{momflux}). To evaluate the momentum-flux term by term, 
let us notice that 
the fourth and the seventh term in (\ref{tilde}), which
are non-zero individually, add up to zero 
asymptotically,
\begin{eqnarray}
{1\over 2h}{\rm e}^{-\sigma}
\rho_{a b}\rho_{c d}\pi^{a c}\pi^{b d} 
+{1\over 8h}{\rm e}^{\sigma}\rho^{a b} \rho^{c d}
(D_{+}\rho_{a c}) (D_{+}\rho_{b d})       
&=&{h\over 2}{\rm e}^{\sigma}\rho^{a b} \rho^{c d}
(D_{-}\rho_{a c}) (D_{-}\rho_{b d})  \nonumber\\
&\longrightarrow &
O\Big(  {1\over v^{2} } \Big),       \label{fourseven}
\end{eqnarray}
where we used the definition (\ref{definition})
of $\pi^{a b}$. All other  non-vanishing terms are given by 
\begin{eqnarray}
& & 
\lim_{v\rightarrow \infty} 
{1 \over 16\pi} \oint_{S_{2}} \!\!\!  d^{2}y \, \Big(
{1\over 4}h{\rm e}^{-\sigma}\pi_{h}^{2} \Big)
={1\over 2},              \label{non1}\\
& &
\lim_{v\rightarrow \infty} 
{1 \over 16\pi} \oint_{S_{2}} \!\! \! d^{2}y \,  \Big(
{1\over 2}{\rm e}^{-\sigma}\pi_{h}\pi_{\sigma} 
\Big)
=1,                     \label{non2}\\
& & 
\lim_{v\rightarrow \infty} 
{1 \over 16\pi} \oint_{S_{2}} \!\! \! d^{2}y \, 
 {\rm e}^{\sigma}R_{2} 
={1\over 4 } \chi,            \label{non3}
\end{eqnarray}
where  $\chi=2$ for a two-sphere $S_{2}$.
Therefore we have 
\begin{equation}
{d \over d u} P_{\rm B}(u)
=0.                      \label{allright}
\end{equation}

\end{subsection}

\begin{subsection}{The Bondi angular momentum and angular 
momentum-flux}

The total angular momentum at the null infinity is naturally 
defined 
as the limiting value of the general quasi-local angular momentum
$L(u,v;\xi)$ in (\ref{goodef}), 
\begin{equation}
\lim_{v\rightarrow \infty} L(u,v;\xi)
:=L_{\rm B}(u;\xi).
\end{equation}
Since the background fields $\bar{n}$ and $\bar{l}$ in (\ref{backn}) 
commute, 
\begin{equation}
[\bar{n}, \ \bar{l} ]_{\rm L}=0,                \label{zerocom}
\end{equation}
it follows that 
\begin{equation}
\bar{L}_{\rm B}(u;\xi)=0                     \label{zerotwi}
\end{equation}
for all $\xi$. Let $\xi$ be asymptotic to the azimuthal Killing vector field
of the Kerr spacetime such that 
\begin{equation}
\xi :=\xi^{a}\partial_{a} 
\longrightarrow {\partial \over \partial \varphi}.
\end{equation}
Then, we have
\begin{eqnarray}
L_{\rm B}(u;\xi) &=&{1\over 16\pi}  \int_{0}^{2\pi} \! \! \! \! \! 
d \varphi \! \int_{0}^{\pi} \! \! \! d \vartheta \, 
(6ma) \, {\rm sin}^{3}\vartheta   \nonumber\\
&=&ma,
\end{eqnarray}
which is just the total angular momentum of the Kerr
spacetime. 

The total angular momentum flux at the instant $u$ is given by the
asymptotically limiting form of the the equation (\ref{angflux}), 
which is 
\begin{equation} 
{d L_{\rm B}\over du}:=\lim_{v\rightarrow \infty}
{1 \over 16\pi}\oint_{S_{2}} \!\! \! d^{2}y \, \Big( 
\pi^{a b}{\pounds}_{\xi} \rho_{a b}  
+\pi_{\sigma}{\pounds}_{\xi} \sigma 
-h {\pounds}_{\xi} \pi_{h} 
-A_{+}^{\ a}{\pounds}_{\xi}\pi_{a} \Big).   \label{angfluxa}
\end{equation} 
Let us evaluate each term in the r.h.s. of this equation. 
The first term is given by
\begin{eqnarray}
\pi^{a b}{\pounds}_{\xi} \rho_{a b} 
&=& \Big\{ -{1 \over 2}
{\rm e}^{\sigma}\rho^{a c}\rho^{b d}(\partial_{+}\rho_{c d}) + O(1) \Big\}
{\pounds}_{\xi} \rho_{a b}   \nonumber\\
&=& -{\rm sin}\vartheta (\partial_{+}C)({\pounds}_{\xi} C)+ O({1\over v}), \
\label{third1}
\end{eqnarray}
so that we have
\begin{equation} 
\oint_{S_{2}} \!\!\! 
d^{2}y \, 
\pi^{a b}{\pounds}_{\xi} \rho_{a b} 
\longrightarrow -\oint_{S_{2}} \!\!\! 
d \Omega \, \, 
(D_{+}C)({\pounds}_{\xi} C).   \label{first}
\end{equation}
The second term becomes
\begin{equation}
\pi_{\sigma}{\pounds}_{\xi} \sigma 
= \Big\{ -2v \, {\rm sin}\vartheta + O(1) \Big\}
{\pounds}_{\xi} \sigma. 
\end{equation}
Since $\sigma$ is asymptotically given by
\begin{equation}
\sigma =2{\rm ln}\, v + {\rm ln}\,|{\rm sin}\vartheta| 
+ {\rm ln}\Big\{  1 + O\Big( {1\over v^{2}} \Big)   \Big\},
\end{equation}
we have
\begin{equation}
{\pounds}_{\xi} \sigma =O\Big( {1\over v^{2}} \Big),
\end{equation}
so that the second term becomes
\begin{equation}  
\oint_{S_{2}} \!\!\! 
d^{2}y \,
\pi_{\sigma}{\pounds}_{\xi} \sigma
=O \Big( {1\over v} \Big) \longrightarrow 0.  \label{second}
\end{equation}
The third term becomes
\begin{eqnarray}
h {\pounds}_{\xi} \pi_{h}
&=& -\pi_{h}  {\pounds}_{\xi} h + {\pounds}_{\xi} (h \pi_{h}) \nonumber\\
&=& {\pounds}_{\xi} (  4 m {\rm sin}\vartheta  +  h \pi_{h}) 
+ O\Big( {1\over v} \Big),  \label{okterm}
\end{eqnarray}
where we used that 
\begin{equation}
{\pounds}_{\xi} ({\rm sin}\vartheta) = 0.
\end{equation}
Thus we have
\begin{equation} 
\oint_{S_{2}} \!\!\! 
d^{2}y \,
h {\pounds}_{\xi} \pi_{h}
=O \Big( {1\over v} \Big) \longrightarrow 0.  \label{third}
\end{equation}
The fourth term is of the order of 
\begin{equation}
A_{+}^{\ a} {\pounds}_{\xi}\pi_{a}
=O \Big( {1\over v^{3}} \Big),
\end{equation}
so that 
\begin{equation} 
\oint_{S_{2}} \!\!\! 
d^{2}y \, 
A_{+}^{\ a} {\pounds}_{\xi}\pi_{a}
=O({1\over v^{3}}) 
\longrightarrow 0.                 \label{fourth}
\end{equation} 
If we put together (\ref{first}), (\ref{second}), (\ref{third}), 
and (\ref{fourth}) into (\ref{angfluxa}), then 
the total angular momentum flux at the null infinity is given by
\begin{equation}
{d L_{\rm B}\over du} 
=-\lim_{v\rightarrow \infty}
{1\over 32\pi}\oint_{S_{2}} \!\!\! 
d^{2}y \,
{\rm e}^{\sigma}\rho^{a c}\rho^{b d}(\partial_{+}\rho_{c d}) 
({\pounds}_{\xi} \rho_{a b}),           \label{aflux}
\end{equation}
or, equivalently,
\begin{equation}
{d L_{\rm B}\over du} 
=-{1\over 16\pi}\oint_{S_{2}} \!\!\! d \Omega \,\,
(\partial_{+}C)({\pounds}_{\xi} C),     \label{totalflux}
\end{equation}
which is precisely the Bondi angular momentum flux at the null 
infinity\cite{ashtekar81}.

\end{subsection}

\begin{subsection}{Gravitational Carter's constant }

Let us define the asymptotic quantity $J_{\rm B}(u)$ as
\begin{equation}
\lim_{v\rightarrow \infty}v^{3}J(u,v):=J_{\rm B}(u).
\end{equation}
Because of the equation (\ref{zerocom}), 
the subtraction term $\bar{J}_{\rm B}(u)$ is zero,
\begin{equation}
\bar{J}_{\rm B}(u)=0.
\end{equation}
Then the asymptotic integral $J_{\rm B}(u)$ becomes
\begin{equation}
J_{\rm B}(u)= 2(ma)^{2}.               \label{jlimit}
\end{equation}
Thus, $J_{\rm B}(u)$  is (twice of) the angular 
momentum squared, deserveing the name the {\it gravitational} 
analog of the Carter's ``fourth'' constant at the null infinity. 

\end{subsection}
\end{section}
\vspace{.7cm}

\begin{section}{In-going null coordinates}

One might be also interested in applying this formalism 
to black holes, 
and try to obtain quasi-local  quantities defined 
on the black hole horizon 
and corresponding fluxes incident on that horizon.
For this problem, it is appropriate to choose a coordinate system
adapted to the {\it in}-going null geodesics.
Such a coordinate system is described by the metric 
\begin{equation}
ds^2 = + 2dudv - 2hdu^2 +{\rm e}^{\sigma} \rho_{ab}
 \left( dy^a + A_{+}^{\ a}du +A_{-}^{\ a} dv \right)  
\left( dy^b + A_{+}^{\ b}du +A_{-}^{\ b} dv 
\right).                  \label{ingoing}
\end{equation}
In this coordinate system, 
the future-directed out-going and in-going null vector fields 
$n'$ and $l'$ are given by
\begin{eqnarray}
& & n':= \hat{\partial}_{+} +  h  \hat{\partial}_{-}, \label{eneh}\\
& & l':= -\hat{\partial}_{-},                   \label{eleh}
\end{eqnarray}
respectively, which are normalized so that
\begin{equation}
< n', \ l'> = - 1.
\end{equation}
The inverse relation is given by
\begin{eqnarray}
& & 
\hat{\partial}_{+}=n' + h l',  \label{ehen}\\
& &
\hat{\partial}_{-}= -l'.  \label {ehel}
\end{eqnarray}
If we repeat the same analysis as in the previous chapters 
using the metric (\ref{ingoing}), 
we obtain a new (but equivalent, of course) set of quasi-local 
conservation equations, which we present without derivations. 
The conjugate momenta are found to be
\begin{eqnarray}
& &\pi_{h}=-2 {\rm e}^{\sigma}D_{-}\sigma,   \label{piheh}\\
& &\pi_{\sigma} = -2 {\rm e}^{\sigma} D_{-}h
       -2h {\rm e}^{\sigma} D_{-}\sigma
     - {\rm e}^{\sigma} D_{+}\sigma,   \label{pisigmaeh}  \\
& &\pi_{a}={\rm e}^{2 \sigma} 
\rho_{a b}F_{+-}^{\ \ b}, \label{piaeh} \\
& &\pi^{a b}= 
h{\rm e}^{\sigma} \rho^{a c}\rho^{b d}D_{-}\rho_{c d}
+{1\over 2}{\rm e}^{\sigma} \rho^{a c}\rho^{b d}
D_{+}\rho_{c d},                            \label{firsteh}
\end{eqnarray}
and the ``Hamiltonian'' function $H$ is given by
\begin{eqnarray}
& & H =-{1\over 2}{\rm e}^{-\sigma}\pi_{h}\pi_{\sigma}   
+ {1\over 4}h{\rm e}^{-\sigma}\pi_{h}^{2}
 -{1\over 2}{\rm e}^{-2\sigma}\rho^{a b}\pi_{a}\pi_{b} 
+{1\over 2h}{\rm e}^{-\sigma}
\rho_{a b}\rho_{c d}\pi^{a c}\pi^{b d} \nonumber\\
& &
-{1\over 2}\pi_{h}D_{+}\sigma  
-{1\over 2h}\pi^{a b}D_{+}\rho_{a b}
+{1\over 8h}{\rm e}^{\sigma}\rho^{a b} \rho^{c d}
(D_{+}\rho_{a c}) (D_{+}\rho_{b d}) 
+{\rm e}^{\sigma}R_{2}.              \label{newham} 
\end{eqnarray}
A new set of quasi-local conservations equations are found to be
\begin{eqnarray}
& &
{\partial \over \partial u} U(u,v)
={1\over 16\pi}  \oint \! d^{2}y \, \Big(
\pi^{a b}  \partial_{+} \rho_{a b} 
+\pi_{\sigma}\partial_{+}\sigma  - h \partial_{+} \pi_{h}
-A_{+}^{\ a} \partial_{+}\pi_{a} \Big),    \label{enfluxeh} \\
& &  {\partial \over \partial u} P(u,v)
=-{1 \over 16\pi} \oint \! d^{2}y \, H,  \label{momfluxeh}  \\
& & {\partial \over \partial u} L(u,v;\xi)
={1 \over 16\pi}\oint \! d^{2}y \, \Big( 
\pi^{a b}{\pounds}_{\xi} \rho_{a b}  
+\pi_{\sigma}{\pounds}_{\xi} \sigma 
-h {\pounds_{\xi}} \pi_{h}   
-A_{+}^{\ a} {\pounds_{\xi}}\pi_{a} \Big)
\hspace{.25in} (\partial_{+}\xi^{a}=0),   \label{angfluxeh}\\ 
& & 
{\partial \over \partial u} J(u,v) 
={1\over 16\pi} 
\oint \! d^{2}y \, \Big( 
\pi^{a b}{\pounds}_{A_{+}} \rho_{a b}  
+\pi_{\sigma}{\pounds}_{A_{+}} \sigma 
-h {\pounds}_{A_{+}} \pi_{h} \Big)  
\hspace{.25in} (\partial_{+}A_{+}^{\ a}=0), \label{jfluxeh}
\end{eqnarray}
where  $U(u,v)$, $P(u,v)$, $L(u,v;\xi)$, and $J(u,v)$ 
are defined as 
\begin{eqnarray}
& & U(u,v):= {1 \over 16\pi}\oint d^{2}y \,   ( 
h\, \pi_{h} - 2 {\rm e}^{\sigma} D_{+}\sigma ) 
+ \bar{U},               \label{eninteh}\\
& &
P(u,v):={1 \over 16\pi} \oint \! d^{2}y \, ( \pi_{h} )
+ \bar{P},                \label{mominteh}\\
& &
L(u,v; \xi):={1 \over 16\pi} \oint \! d^{2}y \, 
(\xi^{a}\pi_{a}) + \bar{L},   \label{aanginteh}\\ 
& & 
J(u,v):= \oint \! d^{2}y \, ( 
A_{+}^{\ a} \pi_{a} ) + \bar{J},             \label{anginteh}
\end{eqnarray} 
where $\bar{U}$, $\bar{P}$, $\bar{L}$, and $\bar{J}$ are
undetermined reference terms as before. 
Notice that we could also have written the equation (\ref{enfluxeh}) as 
\begin{equation}
{\partial \over \partial u} U(u,v)
={1\over 16\pi}  \oint \! d^{2}y \, \Big(
\pi^{a b}  D_{+} \rho_{a b} 
+\pi_{\sigma}D_{+}\sigma  - h D_{+} \pi_{h} \Big), \label{endee}
\end{equation}
as in (\ref{enflux}). 
In geometrical terms, these quasi-local quantities can be expressed
as, 
\begin{eqnarray}
& & U(u,v):= -{1 \over 8\pi} \pounds_{n'}{\cal A}
+{1 \over 8\pi} \pounds_{\bar{n}'}{\cal A},
                              \label{eninteha}\nonumber\\
& &
P(u,v)={1 \over 8\pi}\pounds_{l'}{\cal A} 
-{1 \over 8\pi}\pounds_{\bar{l}'}{\cal A}, \label{mominteha}\nonumber\\
& & 
L(u,v;\xi)
={1 \over 16\pi} \oint \! d^{2}y \, 
{\rm e}^{ \sigma} \xi_{a}  [n, \ l ]_{\rm L}^{a} 
-{1 \over 16\pi} \oint \! d^{2}y \, 
{\rm e}^{ \sigma} \xi_{a}  [\bar{n}', \ \bar{l}']_{\rm L}^{a}
\hspace{.25in} 
(\partial_{+}\xi^{a}=0),         \label{goodeh}\nonumber\\
& &
J(u,v)
={1 \over 16\pi} \oint \! d^{2}y \, 
{\rm e}^{ 2\sigma} \rho_{a b} A_{+}^{\ a}  [n, \ l ]_{\rm L}^{b} 
-{1 \over 16\pi} \oint \! d^{2}y \, 
{\rm e}^{ 2\sigma}  \rho_{a b} A_{+}^{\ a} 
[\bar{n}', \ \bar{l}' ]_{\rm L}^{b}  
\hspace{.25in}
(\partial_{+}A_{+}^{\ a}=0).         \label{goodeha}
\end{eqnarray}
Here $\bar{n}'$, $\bar{l}'$ are future-directed in-going 
and out-going null vector fields of a background reference 
spacetime
\begin{equation}
d\bar{s}^2
= +2dudv - 2\bar{h}du^2 +{\rm e}^{\sigma} \rho_{ab}
 \left( dy^a + \bar{A}_{+}^{\ a}du +\bar{A}_{-}^{\ a} dv \right) 
\left( dy^b + \bar{A}_{+}^{\ b}du +\bar{A}_{-}^{\ b} dv 
\right),                         \label{yoonrefnew}
\end{equation}
such that 
\begin{eqnarray}
& & 
\bar{n}':=\Big( {\partial \over \partial u} 
- \bar{A}_{+}^{\ a}{\partial \over \partial y^{a}} \Big)
+\bar{h}  \Big( 
{\partial \over \partial v} 
- \bar{A}_{-}^{\ a}{\partial \over \partial y^{a}}\Big), \nonumber\\
& & 
\bar{l}':=-  \Big( {\partial \over \partial v} 
- \bar{A}_{-}^{\ a}{\partial \over \partial y^{a}}\Big). \label{enbarnew}
\end{eqnarray}

\end{section}

\begin{section}{Quasi-local horizon}{\label{qhorizon}}

Recall that the event horizon is a global concept which is inseparable 
from the notion of infinity. 
Therefore, in order to discuss the dynamics of black holes 
quasi-locally, we have to introduce a new notion of 
{\it quasi-local horizon}, 
which refers to the quasi-local region only. 
We define the quasi-local horizon ${\cal H}$ 
as a three-dimensional hypersurface
on which the vector field $\hat{\partial}_{+}$, 
which is an arbitrary vector field except that it is asymptotic to 
the timelike Killing vector at the infinity, has a zero norm,
\begin{equation}
< \hat{\partial}_{+}, \hat{\partial}_{+}> 
= -2 h =0.           \label{qsurf1}
\end{equation}
The location of the quasi-local horizon  ${\cal H}$ 
can be found by solving the equation 
\begin{equation}
h(u,v,y^{a})=0                              \label{qsurf}
\end{equation}
for $v$, and 
the generator of the quasi-local horizon is given by
\begin{equation}
\hat{\partial}_{+}= {\partial \over\partial u} 
-A_{+}^{\ a}{\partial \over\partial y^{a}},            \label{gkill}
\end{equation}
which is out-going null on ${\cal H}$.

Let us remark a few properties of this quasi-local horizon. 
First,  the quasi-local horizon is defined for generic
spacetimes that do not have isometries in general, and
its location is not fixed, but {\it varies}
as much as the choice of the vector field 
$\hat{\partial}_{+}$ does. 
In this sense the quasi-local horizon is not a spacetime invariant
but a {\it covariant} notion.
Even the signature of the vector field 
$\hat{\partial}_{+}$ is not determined {\it a priori}. 
In the region where $\hat{\partial}_{+}$ is non-spacelike, however,
the quasi-local horizon may be regarded as a generalization of 
the Killing horizon, since it is defined as the hypersurface where 
$\hat{\partial}_{+}$ has a zero norm.
For instance, 
for the Schwarzschild solution, we have
\begin{equation}
2h=1-{2m \over v},
\end{equation}
so that $h=0$ on the Killing horizon $v=2m$. 
Moreover, since the quasi-local horizon is generated by the
out-going null vector fields, all the fluxes
crossing the quasi-local horizon is purely {\it in-going}, just like
the stretched horizon of Price and Thorne.

In this section, we shall delimit our discussions of the quasi-local 
conservation equations to the quasi-local 
horizon ${\cal H}$, and find that the quasi-local
conservation equations on ${\cal H}$ coincides {\it exactly} 
with the quasi-local conservation equations of Price and Thorne\cite{pt86}
defined on the {\it stretched} horizon. 

\begin{subsection}{Surface gravity {$\kappa$}} 

In order to discuss the dynamics of quasi-local horizon, 
it is useful to introduce the notion of the surface gravity $\kappa$ 
to the generic, time-dependent, quasi-local horizon. 
On the quasi-local horizon ${\cal H}$ on which $h=0$, 
the vector field  $\chi$ defined as
\begin{equation}
\chi:={\partial \over \partial u} 
- A_{+}^{\ a}{\partial \over \partial y^{a}}  \label{chig}
\end{equation}
becomes a generator of ${\cal H}$, since we have 
\begin{equation}
\chi\cdot \chi  =-2h  |_{\cal H} = 0. 
\end{equation}
Hence $\nabla_{A}(\chi\cdot \chi)|_{\cal H}$
is normal to ${\cal H}$, 
which means that there exists a function $\kappa$ 
defined on ${\cal H}$ such that
\begin{equation}
\nabla_{A}(\chi\cdot \chi)|_{\cal H}
:=-2\kappa \chi_{A}|_{\cal H},
\hspace{.2in}                 \label{surf}
\end{equation}
where $A$ is a spacetime index such that $A=\{ +,- ,a \}$. 
Notice that this function $\kappa$ can be defined
on any null hypersurface. When the null hypersurface coincides 
with the event horizon, this function is the {\it surface gravity} of 
the corresponding black hole. But one may use the 
same terminology for $\kappa$ on a quasi-local horizon ${\cal H}$,
since it is a (segment of) null hypersurface 
beyond which a local observer whose worldline $\hat{\partial}_{+}$
has a zero norm on ${\cal H}$ does not have access to.
It may be instructive to notice that, 
for the Kerr black hole, $\chi$ is the generator of 
the event horizon, since it becomes
\begin{equation}
\chi \longrightarrow  {\partial \over \partial u} 
- \Omega_{\cal H}{\partial \over \partial \varphi},  \label{chikerr}
\end{equation}
where ${\partial / \partial u}$ and
${\partial / \partial \varphi}$ are timelike and axial
Killing vector fields, and $\Omega_{\cal H}$ is 
the angular velocity of the Kerr horizon relative to the Killing time.

Let us compute $\kappa$ on the quasi-local horiozn ${\cal H}$.
In the basis $\{ {\partial / \partial u}$, 
${\partial / \partial v}$, ${\partial / \partial y^{a}} \} $,
the components of $\chi$ are given by
\begin{eqnarray}
& & \chi^{A}= (1,\ 0,\ -A_{+}^{ \ a}),   \nonumber\\
& & \chi_{A} :=g_{AB}\chi{^B} = (-2h,\ 1,\ 0).       \label{concov}
\end{eqnarray}
If we put (\ref{concov}) into (\ref{surf}), then we find that
\begin{equation}
\partial_{+}h|_{\cal H} 
= \partial_{a}h|_{\cal H} = 0,
\end{equation} 
and that $\kappa$ is given by
\begin{equation}
\kappa=D_{-}h|_{\cal H}.           \label{dersur}
\end{equation}
Notice that, in general, $\kappa$ is {\it not} constant over 
${\cal H}$ so that 
\begin{equation}
\partial_{+}\kappa \neq 0,    \hspace{.2in}
\partial_{a}\kappa \neq 0,
\end{equation}
which reflects the dynamical nature of the quasi-local horizon 
${\cal H}$.

\end{subsection}

\begin{subsection}{Quasi-local energy conservation on ${\cal H}$}

Notice that if we restrict the energy equation (\ref{endee}) to 
the quasi-local horizon ${\cal H}$, then it becomes, 
\begin{eqnarray}
& & {\partial \over \partial u}U_{\cal H}
={1\over 16\pi}  \oint_{\cal H} \! d^{2}y \, \Big\{
{1\over 2} {\rm e}^{\sigma} \rho^{a b}\rho^{c d}
(D_{+}\rho_{a c}) (D_{+}\rho_{b d})  
-{\rm e}^{\sigma} (D_{+}\sigma)^2  
  -2 {\rm e}^{\sigma} \kappa D_{+}\sigma
   \Big\},              \label{enfluxa} \\
& & U_{\cal H}:= -{1 \over 8\pi}
\oint_{\cal H} d^{2}y \,    
{\rm e}^{\sigma} D_{+}\sigma 
+ \bar{U}_{\cal H}.               \label{enintb}
\end{eqnarray}
This equation is identical to the integral of the following 
equation
\begin{equation}
{\partial \over \partial u} \Sigma_{\cal H} 
+ \theta_{\cal H} \Sigma_{\cal H} 
=-{1\over 8\pi}\kappa\ \theta_{\cal H}   
-{1\over 16\pi} \theta_{\cal H}^{\ 2} 
+ {1\over 8\pi} \sigma_{a b}^{\cal H} \sigma_{\cal H}^{a b}
\end{equation}
over the stretched horizon of Price and Thorne, 
where the notations are such that 
\begin{eqnarray}
& &
\Sigma_{\cal H}=- {1\over 8\pi}\theta_{\cal H}, \nonumber\\
& & 
\theta_{\cal H}=D_{+}\sigma, \nonumber\\
& & 
\sigma_{a b}^{\cal H}
={1\over 2}{\rm e}^{\sigma}D_{+}\rho_{a b},  \nonumber\\
& & 
\sigma_{{\cal H} }^{ a b} 
=\phi^{a c}\phi^{b d}\sigma_{ c d}^{\cal H} 
={1\over 2}{\rm e}^{-\sigma}\rho^{a c}\rho^{b d}D_{+}\rho_{c d}, \nonumber\\
& &
\kappa=D_{-}h|_{\cal H}.                          \label{expan}
\end{eqnarray}
This equation was studied in detail in Eq. (6.112,E) in \cite{pt86}.

It is interesting to discuss the limiting case
when the quasi-local horizon ${\cal H}$ coincides with the event 
horizon. When this happens, the area $A_{\cal H}$ of ${\cal H}$ always 
increases due to the area theorem, so that we have
\begin{equation} 
{d A_{\cal H} \over d u} 
= \oint_{\cal H} \! d^{2}y \, 
( {\rm e}^{\sigma} D_{+}\sigma ) 
\geq 0.
\end{equation}
Furthermore, if the subtraction term 
$\bar{U}_{\cal H}$ is chosen zero, then by the equation (\ref{enintb}), 
$U_{\cal H}$ is non-positive, and when the black hole no longer expands so that
$D_{+}\sigma|_{\cal H}=0$, then
$U_{\cal H}$ becomes zero. For instance, 
for a Schwarzschild or Kerr black hole\cite{pt86}, we have 
\begin{equation}
U_{\cal H}=0 \hspace{.2in} {\rm if} \ \ \bar{U}_{\cal H}\equiv 0.
\end{equation}  
This counter-intuitive 
aspect is a manifestation of the well-known teleological nature of 
the event horizon. That is, when the event horizon evolves, 
its quasi-local energy must be negative so as to cancel 
out the positive in-flux of energy carried by subsequently 
in-falling matter or gravitational radiation, 
leaving  $U_{\cal H}=0$ when the black hole 
reaches the final stationary state. 

\end{subsection}

\begin{subsection}{Quasi-local momentum conservation 
equation on ${\cal H}$} 

Let us mention that the momentum 
equation (\ref{momfluxeh}) 
has a similar structure to the integrated Navier-Stokes equation
for a viscous fluid\cite{lan-lif6},
\begin{equation}
{\partial P_{i} \over \partial u} 
= - \oint \!d S^{k} \Big(   
p\delta_{ik} +\rho v_{i}v_{k} -\sigma'_{ik}  \Big), \label{navier}
\end{equation}
where $ P_{i}$ and $\sigma'_{ik}$ are the total momentum
and the viscous term, 
\begin{eqnarray}
& & P_{i} = \int \! d V \, (\rho v_{i}),   \label{euler1}\\
& & \sigma'_{ik} =\eta\Big(  
{\partial v_{i} \over \partial x^{k}} 
+{\partial v_{k} \over \partial x^{i}} 
-{2\over 3}\delta_{ik}{\partial v_{l} \over \partial x^{l}} \Big)
+\zeta\delta_{ik}{\partial v_{l} \over \partial x^{l}}, 
                              \label{euler2}
\end{eqnarray}
and $\eta$ and $\zeta$ are the coefficients of 
shear and bulk viscosity, respectively. 
This equation tells us that the rate of the net momentum change 
of a fluid within a given volume is determined by
the net momentum-flux across the 
two-surface enclosing the volume. Notice that
the ``Hamiltonian'' function $H$ in (\ref{newham}), which is 
at most quadratic in the conjugate momenta $\pi_{I}$, 
assumes the form of momentum-flux of a viscous fluid. 
Namely, terms quadratic in $\pi_{I}$ may be viewed as
responsible for direct momentum transfer, 
terms linear in $\pi_{I}$ as viscosity terms, 
and terms independent of $\pi_{I}$ as pressure terms. 
From this point of view, one may interpret the ``Hamiltonian'' function
$H$ as the gravitational momentum-flux and 
the two-surface integral
\begin{equation}
P_{\cal H}={1 \over 16\pi} \oint \! d^{2}y \, ( \pi_{h} ) 
+ \bar{P}                \label{momintz}
\end{equation}
as the quasi-local gravitational momentum within $N_{2}$. 
On the quasi-local horizon ${\cal H}$, the equation 
(\ref{momfluxeh}) becomes, 
\begin{eqnarray}
& & {\partial \over \partial u}P_{\cal H}
=-{1\over 16\pi}  \oint_{\cal H} \! d^{2}y \, \Big(
{\rm e}^{\sigma}R_{2}
-{1\over 2}{\rm e}^{-2 \sigma}\rho^{a b}
\pi_{a}\pi_{b}
-{1\over 2}{\rm e}^{- \sigma}\pi_{\sigma}\pi_{h}
-{1\over 2}\pi_{h} D_{+}\sigma \Big).  \label{momfluxb}
\end{eqnarray}
The first term on the r.h.s. is given by the Euler number
$\chi$, 
\begin{equation}
{1 \over 16\pi} \oint_{\cal H} \!\! \! d^{2}y \, 
{\rm e}^{\sigma}R_{2}
={1\over 4 } \chi,                   \label{non33}
\end{equation}
where  $\chi=2$ for a two-sphere. The second and third terms are 
quadratic in the momenta, and the last term is linear in the 
momentum.
It is curious that this conservation equation 
is missing in the work of Price and Thorne\cite{pt86}. 
Let us note that, in terms of the configuration variables, 
the integrand of the r.h.s. of (\ref{momfluxb}) can be written as
\begin{equation}
H_{\cal H} = {\rm e}^{\sigma}R_{2}   
-{1\over 2}{\rm e}^{2 \sigma}\rho_{a b}
F_{+-}^{\ \ a}F_{+-}^{\ \ b}
-2 {\rm e}^{\sigma}\kappa D_{-}\sigma.   \label{hameh} 
\end{equation}

\end{subsection}
\begin{subsection}{Quasi-local angular momentum conservation 
equation on ${\cal H}$}

In this section, we shall show that the equation 
(\ref{angfluxeh}) when restricted to the surface ${\cal H}$
coincides with the quasi-local angular momentum equation of Price and Thorne
on the stretched horizon. 
Let us first notice that the equation (\ref{angfluxeh}) 
on ${\cal H}$ can be written as
\begin{eqnarray}
& & {\partial \over \partial u}L_{\cal H}
={1\over 16\pi}  \oint_{{\cal H}} \! d^{2}y \, \Big\{
-{\rm e}^{\sigma}  (2 \kappa +D_{+}\sigma) {\pounds_{\xi}}\sigma
+{1\over 2} {\rm e}^{\sigma} \rho^{a c}\rho^{b d}
(D_{+}\rho_{a b})({\pounds}_{\xi}\rho_{c d})
-A_{+}^{ \ a}{\pounds_{\xi}}\pi_{a}  \Big\},       \label{angfnew} \\
& & L_{\cal H}:= {1 \over 16\pi}
\oint_{\cal H} d^{2}y \,  ( \xi^{a}\pi_{a} )
+ \bar{L}_{\cal H}.               \label{angm}
\end{eqnarray}
The equation (\ref{angfnew}) turns out to be identical 
to the angular momentum conservation equation of Price and Thorne, 
which is given by
\begin{eqnarray}
& & 
D_{+}\Pi^{\cal H}_{ a} 
+ \theta_{\cal H}\Pi^{\cal H}_{ a} 
=-{1\over 8\pi} \ \kappa_{,a}   
-{1\over 16\pi} \theta_{{\cal H}, a} 
+ {1\over 8\pi} \sigma_{{\cal H} a \ ;b}^{\ \ \ b}, 
                              \label{thornang}\\
& & \Pi_{a}^{\cal H}:= -{1\over 16\pi}
{\rm e}^{\sigma}\rho_{a b}F_{+-}^{\ \ b}.  \label{thornem}
\end{eqnarray}
To show this, let us write the equations (\ref{angfnew}) and (\ref{angm})
as 
\begin{eqnarray}
& & {\partial \over \partial u}L_{\cal H}
={1\over 16\pi}  \oint_{{\cal H}} \! d^{2}y \, \Big\{
2{\rm e}^{\sigma} {\pounds_{\xi}}\kappa
+{\rm e}^{\sigma} {\pounds_{\xi}}(D_{+}\sigma) 
+{1\over 2} {\rm e}^{\sigma} \rho^{a b}
{\pounds}_{\xi}(D_{+}\rho_{a b})
+{\rm e}^{2\sigma}\rho_{a b}F_{+-}^{\ \ a}{\pounds_{\xi}}
A_{+}^{ \ b} \Big\},              \label{angfluxb} \\
& & L_{\cal H}:= {1 \over 16\pi}
\oint_{\cal H} d^{2}y \,    
({\rm e}^{2\sigma}\rho_{a b} F_{+-}^{\ \ a}\xi^{b})
+ \bar{L}_{\cal H},               \label{anginteg}
\end{eqnarray}
where we used the identity
\begin{equation}
\oint_{{\cal H}} \! d^{2}y \,  {\pounds_{\xi}} f 
=\oint_{{\cal H}} \! d^{2}y \,  \partial_{a} (\xi^{a}f)
=0
\end{equation}
for any scalar density $f$ with the weight $-1$. 
Using the definitions of $L_{\cal H}$ in (\ref{anginteg})
and $\Pi_{a}^{\cal H}$ in (\ref{thornem}), 
we obtain the following identity,
\begin{eqnarray}
& & 
{\partial \over \partial u}L_{\cal H} 
-{1\over 16\pi}\oint_{{\cal H}} \! d^{2}y \,
( {\rm e}^{2\sigma}\rho_{a b}F_{+-}^{\ \ a}{\pounds_{\xi}} 
A_{+}^{ \ b} )             \nonumber\\
&= & 
-\oint_{{\cal H}} \! d^{2}y \,  \Big\{
{\rm e}^{\sigma} \xi^{a} (D_{+}\sigma) \Pi_{a}^{\cal H}
+{\rm e}^{\sigma}(D_{+}\xi^{a})\Pi_{a}^{\cal H}
+{\rm e}^{\sigma}\xi^{a}D_{+}\Pi_{a}^{\cal H}
+{\rm e}^{\sigma}\Pi_{a}^{\cal H}[A_{+}, \ \xi]_{\rm L}^{a} 
       \Big\}                              \nonumber\\
&= &
-\oint_{{\cal H}} \! d^{2}y \,\Big\{
{\rm e}^{\sigma}\xi^{a} D_{+}\Pi_{a}^{\cal H}
+{\rm e}^{\sigma}\xi^{a}(D_{+}\sigma) \Pi_{a}^{\cal H}
+{\rm e}^{\sigma}\Pi_{a}^{\cal H} \Big( D_{+}\xi^{a}
+[A_{+}, \ \xi]_{\rm L}^{a} \Big)
\Big\}  \nonumber\\
&= &
-\oint_{{\cal H}} \! d^{2}y \,
{\rm e}^{\sigma}\xi^{a}(
D_{+}\Pi_{a}^{\cal H} +  \theta_{\cal H}\Pi_{a}^{\cal H}),    \label{ang1}
\end{eqnarray}
where we used the diff$N_{2}$-covariant 
derivative of $\xi^{a}$, 
\begin{equation}
D_{+}\xi^{a}:=\partial_{+}\xi^{a} 
- [A_{+}, \ \xi]_{\rm L}^{a}
\end{equation}
defined in the section \ref{intro}, 
and the condition that  
\begin{equation}
\partial_{+}\xi^{a}=0.              \label{xicon1}
\end{equation}
The first and second term on the r.h.s. of 
(\ref{angfluxb}) become, 
\begin{equation}
{1\over 16\pi}\oint_{{\cal H}} \! d^{2}y \,\Big\{
2{\rm e}^{\sigma} {\pounds_{\xi}}\kappa
+{\rm e}^{\sigma} {\pounds_{\xi}}(D_{+}\sigma) \Big\}
=\oint_{{\cal H}} \! d^{2}y \,
{\rm e}^{\sigma}\xi^{a}
\Big(
{1\over 8\pi}\kappa,_{a} +{1\over 16\pi}\theta_{{\cal H}, a} 
\Big),                                \label{ang2}
\end{equation}
and the third term on the r.h.s. of (\ref{angfluxb}) is given by
\begin{eqnarray}
{1\over 2} {\rm e}^{\sigma} \rho^{a b}
{\pounds}_{\xi}(D_{+}\rho_{a b})  
&=&
{1\over 2} {\rm e}^{\sigma} \rho^{a b} \Big\{
\xi^{c} \nabla_{c} (D_{+}\rho_{a b}) 
+(D_{+}\rho_{c b})(\nabla_{a} \xi^{c})
+(D_{+}\rho_{a c})(\nabla_{b} \xi^{c})
-(\nabla_{c} \xi^{c})(D_{+}\rho_{a b})   \Big\}   \nonumber\\
&=&
{1\over 2} {\rm e}^{\sigma}\xi^{c} 
\Big\{ 
\nabla_{c} ( \rho^{ab} D_{+}\rho_{a b} )
-(\nabla_{c} \rho^{ab} )D_{+}\rho_{a b} \Big\}
+ {\rm e}^{\sigma} \rho^{a b} (D_{+}\rho_{b c}) (\nabla_{a}\xi^{c})
                      \nonumber\\
&=&{\rm e}^{\sigma} \rho^{a b} (D_{+}\rho_{b c}) 
(\nabla_{a}\xi^{c})                         \nonumber\\
&=& 
-\xi^{a}{\rm e}^{\sigma}  \rho^{b c} \nabla_{b}(D_{+}\rho_{a c})
+{\rm e}^{\sigma} \rho^{a b} 
\nabla_{a} (\xi^{c}D_{+}\rho_{b c})             \nonumber\\
&=&
-2\xi^{a} {\rm e}^{\sigma}\sigma_{{\cal H} a \ ;b}^{\ \ \ b}
+ \nabla_{a} (\xi^{c}{\rm e}^{\sigma} 
\rho^{a b}D_{+}\rho_{b c}).
\end{eqnarray}
Here we used the unimodular condition (\ref{module}), 
and the metricity condition
\begin{equation}
\nabla_{a}\sigma = \nabla_{a}\rho_{b c}=0.
\end{equation}
Notice that the shear tensor $\sigma_{{\cal H} a }^{ \ \ \ b}$
is given by
\begin{equation}
\sigma_{{\cal H} a }^{ \ \ \ b}
:=\phi^{b c}\sigma_{{\cal H}a c}={1\over 2}\rho^{b c}D_{+}\rho_{a c}.
\end{equation}
Therefore we have 
\begin{equation}
\oint_{{\cal H}} \! d^{2}y \,
{1\over 2} {\rm e}^{\sigma} \rho^{a b}
{\pounds}_{\xi}(D_{+}\rho_{a b})   
=
-\oint_{{\cal H}} \! d^{2}y \,
2\xi^{a} {\rm e}^{\sigma}\sigma_{{\cal H} a \ ;b}^{\ \ \ b}.  
                                       \label{ang3}
\end{equation}
If we put together (\ref{ang1}), (\ref{ang2}), 
and (\ref{ang3}), then (\ref{angfluxb}) becomes
\begin{equation}
\oint_{{\cal H}} \! d^{2}y \,  
{\rm e}^{\sigma}\xi^{a}\Big\{
D_{+}\Pi_{a}^{\cal H} + \theta_{\cal H} \Pi_{a}^{\cal H}
+{1\over 8\pi}\kappa,_{a} +{1\over 16\pi}\theta_{{\cal H}, a} 
-{1\over 8\pi}
\sigma_{{\cal H} a \ ;b}^{\ \ \ b} \Big\} =0,
\end{equation}
for an arbitrary function $\xi^{a}$ that satisfies (\ref{xicon1}).
This shows that the two equations (\ref{angfnew}) and (\ref{thornang}) 
are identical.

\end{subsection}

\end{section}

\begin{section}{Summary}
\label{discuss}

In this note, I presented a set of quasi-local 
conservation equations that were found while studying
the Einstein's equations using the (1+1)-dimensional description, 
and studied 
physical significances of these conservation equations.
The key observation is that the affine coordinate $v$ can be treated 
as a natural time coordinate in this formalism.
The subsequent Hamiltonian formalism was developed with respect 
to this time coordinate. One of the outcomes
of this analysis is the discovery of a set of Bondi-like 
quasi-local conservation equations for the vacuum general 
relativity, which reproduce {\it both}
the well-known conservation 
equations in the asymptotic null infinity in asymptotically flat spacetimes
{\it and} the corresponding conservation equations on the inner 
quasi-local horizon 
of a generic dynamical spacetime.
All of these quasi-local quantities are expressed in geometrically
invariant terms such as the area of the two-surface and
a pair of null vector fields orthogonal to that surface.
It was also found that each quantity has a natural interpretation 
as the quasi-local energy, linear momentum, and angular momentum 
of a two-surface and corresponding fluxes crossing that surface. 

In addition to the above quasi-local quantities, we also obtained the 
quasi-local analog of the Carter's ``fourth''
constant of {\it gravitational} field, 
which is somewhat like the angular momentum squared,
measuring the ``intrinsic'' angular momentum of a two-surface. 
The Carter's ``fourth'' constant is known to exist for 
spacetimes that possess two commmuting Killing vector fields 
such as the Kerr black hole.
But in our analysis, it was found that the quasi-local analog of 
Carter's constant exists under the condition
\begin{equation}
\partial_{+}A_{+}^{\ a}=0,
\end{equation}
which is much less restrictive than the existence of two-commuting 
Killing fields. 

Dynamics of the quasi-local horizon was discussed briefly, but 
it deserves further studies. Applications of this formalism to 
astrophysical problems involving black holes and gravitational radiations 
are extremely challenging. These problems are left for future works.

\end{section}

\bigskip\noindent       
\centerline{\bf Acknowledgments}\\

This work was supported by Konkuk University research grant in 2002.

\nopagebreak

\end{document}